\documentclass[useAMS,usenatbib,usegraphicx]{mn2e}

\title[Sodium overabundance of the Hyades giants]{On the sodium overabundance of giants in open clusters: The case of the Hyades\thanks{Based 
on data obtained from the ESO Science Archive Facility. The observations were made with ESO Telescopes at the La Silla Paranal 
Observatory under programmes ID 070.D-0421, 072.C-0393, and 083.A-9011.}}
\author[R. Smiljanic]{Rodolfo Smiljanic\thanks{E-mail:
rsmiljan@eso.org} \\
European Southern Observatory, Karl-Schwarzschild-Str. 2, 85748 Garching bei M\"unchen, Germany}
\begin{document}

\date{Accepted . Received ; in original form }

\pagerange{\pageref{firstpage}--\pageref{lastpage}} \pubyear{2011}

\maketitle

\label{firstpage}

\begin{abstract}
Sodium abundances have been determined in a large number of giants of open clusters but conflicting results, 
ranging from solar values to overabundances of up to five orders of magnitude, have been found. The reasons for this 
disagreement are not well-understood. As these Na overabundances can be the result of deep mixing, their   
proper understanding has consequences for models of stellar evolution. As discussed in the literature, part of this 
disagreement comes from the adoption of different corrections for non-LTE effects and from the use of different atomic data for the 
same set of lines. However, a clear picture of the Na behaviour in giants is still missing. To 
contribute in this direction, this work presents a careful redetermination of the Na abundances of the Hyades giants, motivated 
by the recent measurement of their angular diameters. An average of [Na/Fe] = +0.30, in NLTE, has been found. This overabundance can be 
explained by hydrodynamical models with high initial rotation velocities. This result, and a trend of increasing Na 
with increasing stellar mass found in a previous work, suggests that there is no strong evidence of Na overabundances in red giants beyond those 
values expected by evolutionary models of stars with more than $\sim$ 2 M$_{\odot}$.

\end{abstract}

\begin{keywords}
open clusters and associations: individual: Hyades -- stars: abundances -- 
stars: evolution -- stars: fundamental parameters.
\end{keywords}

\section{Introduction}

In many stages of their evolution, low- and intermediate-mass stars show signs of mixing between 
material of the surface with material of the interior that has been processed by nuclear 
reactions \citep[][and references therein]{Pin97,ipCT08,Sm09b}.

The standard model of stellar evolution, where convection is the only mixing mechanism, does not 
account for all the observational details. The introduction of non-standard physical 
processes, such as atomic diffusion, rotation-induced mixing, internal 
gravity waves, magnetic buoyancy and thermohaline mixing, is unavoidable 
\citep[see e.g.][and references therein]{MoSc00,Young03,TC05,PC06,DePi08,Den09,CharbonnelLagarde10,Mi04,Michaud10,Angelou11,Palmerini11}.

As a star leaves the main sequence towards the red giant branch (RGB), its convective envelope deepens, causing first a 
dilution of lithium, beryllium, and boron \citep{Leb99,PRZ04,Sm10,CantoMartins11} and then the 
first dredge-up \citep{I67}, when material affected by hydrogen burning is mixed to the surface. The dredge-up causes an increase of the abundance 
of nitrogen and a decrease of carbon and of the $^{12}$C/$^{13}$C ratio \citep{CBW98,Gr00}.

An extensive literature has shown a further modification of the surface abundances after the bump 
in the luminosity function on the RGB. At this phase the abundances of Li, C and the ratio $^{12}$C/$^{13}$C are further decreased 
while that of N increases. This effect has been detected in stars of the field and of both open and globular clusters \citep[see e.g.][and references therein]{SPVB86,G89,GB91,CBW98,Gr00,T00,T05,Tautvaisiene10,Smith02,P03,Pi03,Gei05,Sp06,RBdL07,Sm09b,Mikolaitis10,Mikolaitis11,Suda11}.

\subsection{The case of sodium}

During the first dredge-up, Na produced by the NeNa-cycle \citep{DD90} in H-burning regions can 
potentially be mixed to the surface. The observational behaviour of this element, however, is not clear. In open clusters, 
stars with Na 
overabundances as high as [Na/Fe]\footnote{[A/B] = log [N(A)/N(B)]$_{\rm \star}$ $-$ log
 [N(A)/N(B)]$_{\rm\odot}$} = +0.50 or more have been reported \citep{Brag01,Fri03,Jac07,Schuler09}, 
while others were found to have mild overabundances of [Na/Fe] $\sim$ +0.20 \citep{Ham00,T00,PRZ04,Friel10} 
or abundances close to solar \citep{R06,Ses07,Sm09b,Pancino10}.


One of the factors behind this discrepancy is the adoption of different log $gfs$ for the same Na lines by different authors. For example, 
the log $gfs$ adopted in \citet{Sm09b}, from the NIST database \citep{NIST10}, are on average 0.22 dex higher than those 
adopted by \citet{Jac07}, derived with respect to Arcturus. Indeed, as discussed later by \citet{Jac08}, a revision of their $gf$-values 
resulted in an increase of about 0.20 dex with corresponding decrease in the Na abundances. The use of this revised scale now results in 
modest overabundances, [Na/Fe] = +0.10...+0.20 \citep[see Table 14 of][]{Friel10} in comparison with the strong overabundances 
found in \citet{Jac07}, [Na/Fe] = +0.40...+0.60.

Another issue are departures from the local thermodynamic equilibrium (LTE). Several authors provide non-LTE 
corrections for Na \citep{BBG98,GCEG99,MSS00,Tak03,SGZ04,And07b,Lind11b}. As discussed by 
\citet{Lind11b}, there is a scatter of  0.10 to 0.20 dex among the different corrections. Usually non-LTE 
abundances are smaller than the LTE ones. 

In the abundance analyses cited above, the different authors made different choices regarding these matters. It is 
thus difficult to make sense out of these results and understand whether there is a real problem with 
the sodium overabundances in giants. Dedicated studies aiming to better understand this issue are 
still lacking in the literature.

The re-analysis of Na in the Hyades giants presented here is a step on this direction. It is motivated by the recent determination of 
angular diameters (and thus fundamental effective temperatures -- T$_{\rm eff}$) for the four Hyades giants from interferometric measurements using the CHARA array by \citet{Boyajian09}. Accurate Na abundances can help clarifying whether the strong overabundances are real and test whether there is an offset between observations and 
evolutionary models.

This paper is divided as follows. Section \ref{sec:hya} presents the observational data used in the analysis and the determination of the 
atmospheric parameters of the Hyades giants. Section \ref{sec:abun} presents the 
determination of Na abundance from equivalent widths and spectrum synthesis while Sec. \ref{sec:dis} presents a discussion 
of the results. Section \ref{sec:con} summarizes the conclusions.

\section{The Hyades}\label{sec:hya}

The Hyades are the closest open cluster to the Sun \citep[$\sim$ 46.5 pc,][]{VanLeeuwen09}. They have an age of
 625 $\pm$ 50 Myr \citep{Perryman98} and a metallicity of [Fe/H] = +0.13 $\pm$ 0.01 \citep{Paulson03}. There is 
 no sign of Na overabundance in the dwarf stars, i.e. [Na/Fe] $\sim$ 0.00 \citep{Paulson03}. For the giants an average 
 of [Na/Fe] = +0.48 (in LTE) was found by \citet{Schuler09}, which the authors regard as ``too large to be explained by 
 any known self-enrichment scenario''. 

The cluster has four evolved members, all He-burning clump giants: $\epsilon$ Tau (HR 1409 or HD28305), $\gamma$ Tau (HR 1346 or 
HD27371),  $\delta^{1}$ Tau (HR 1373 or HD 27697), and $\theta^{1}$ Tau (HR 1411 or HD 28307). 
A fifth suggested giant member, $\delta$ Ari (HR 951) is likely a non-member \citep{deBruijne01}. The stars $\epsilon$ Tau 
and $\gamma$ Tau are single stars while $\delta^{1}$ Tau and $\theta^{1}$ are spectroscopic binaries \citep{Griffin88,MeMU08}. In 
addition, the star $\epsilon$ Tau was found to have a massive planetary companion \citep{Sato07}.

\begin{figure}
\centering
\includegraphics[width=65mm]{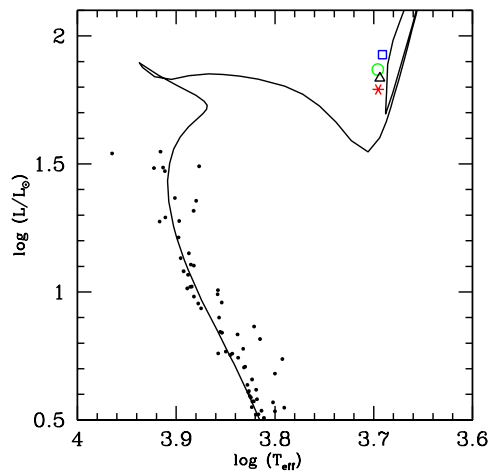}
\caption{HR diagram of the Hyades on the region around the turn-off and the clump. The four giants 
are shown with different symbols ($\epsilon$ Tau as the blue square, $\gamma$ Tau as the green circle, $\delta^{1}$ Tau as 
the black triangle, and $\theta^{1}$ Tau as the red starred symbol). Luminosities and temperatures are from \citet{deBruijne01}.}
\label{fig:hr}
\end{figure}

An HR diagram of the Hyades on the region around the turn-off and the clump is show in Fig. \ref{fig:hr}. An isochrone from \citet{Gir02} with 
625 Myr and [Fe/H] = +0.13 indicates that a clump giant in the Hyades has $\sim$ 2.48 M$_{\odot}$. To illustrate, the corresponding isochrone is also 
shown in Fig. \ref{fig:hr} (no attempt to find the best fitting model was made). 

\subsection{Observational data}

High-resolution spectra of three Hyades giants obtained with FEROS 
\citep{Kau99} at the 2.2m MPG/ESO telescope at La Silla and with UVES \citep{ipDe00} fed by the UT2 of the VLT at Paranal are analyzed here.

FEROS is a fiber-fed echelle spectrograph that provides a full wavelength coverage of $\lambda$$\lambda$350$-$920 nm over 39 orders at
R = 48\,000. The spectra were reduced using the FEROS Data Reduction System (DRS) as implemented within ESO-MIDAS. 
UVES is a cross-dispersed echelle spectrograph able to obtain spectra from the atmospheric cut-off at 300 nm to $\sim$ 1100 nm. Reduction 
was done with the ESO UVES pipeline within MIDAS \citep{uvespipeline}. 

The FEROS data of stars $\delta^{1}$ Tau and $\gamma$ Tau were made available to the author by Luca Pasquini (2010, private 
communication).  The FEROS spectra of $\epsilon$ Tau and the UVES spectra of $\epsilon$ Tau and $\delta^{1}$ Tau were 
retrieved from the ESO/ST-ECF science archive facility. The log book of the observations is given in Table \ref{tab:log}.

%
\begin{table*}
\caption{Observational data.} \label{tab:log} 
\centering
\begin{tabular}{llcccc}
\hline
Star & Spectrograph & $V$ & Date of          & Exp. Time   &  S/N  \\
   &          &            &  observation &  (s)                &  @ 617 nm   \\
\hline
$\gamma$ Tau   & FEROS & 3.654 & 05 Mar. 2004 & 120 &   500 \\
$\epsilon$ Tau   & FEROS & 3.540  & 03 Oct. 2009 & 180 & 500  \\
   & UVES RED 580  &             & 30 Nov. 2002 & 2 $\times$ 1 &  400 \\
$\delta^{1}$ Tau & FEROS & 3.764 &  31 Oct. 2000 & 120 & 700  \\
 & UVES RED 580 &            & 30 Nov. 2002 & 2 $\times$ 1 & 250 \\
\hline
\end{tabular}
\end{table*}
\begin{table*}
\caption{Effective temperatures of the Hyades giants taken from selected recent works from the literature.} \label{tab:atm}
\centering
\begin{tabular}{cccc}
\hline
Star & T$_{\rmn{ eff}}$ & Method & Reference \\
         &   (K)   &     &  \\
\hline
$\gamma$ Tau & 4844 $\pm$ 47 & Interferometry & \citet{Boyajian09} \\
                             & 4965 $\pm$ 75 &  IRFM$^{1}$ & \citet{Schuler09} \\
                             & 4960 $\pm$ 8.1& Line-depth ratio (average) &  \citet{Kovtyukh06} \\
                             & 4981 $\pm$ 80 & Average & Selected results from the PASTEL catalogue \\
$\delta^{1}$ Tau & 4826 $\pm$ 51 & Interferometry & \citet{Boyajian09} \\
                             & 4938 $\pm$ 75 & IRFM$^{1}$ &  \citet{Schuler09} \\
                              & 4975 $\pm$ 7.6 & Line-depth ratio  (average)&  \citet{Kovtyukh06} \\
                              & 5000 $\pm$ 80 & FeI excitation equilibrium &  \citet{HeMe07} \\
                              & 4968 $\pm$ 82 & Average  & Selected results from the PASTEL catalogue \\
$\epsilon$ Tau & 4827 $\pm$ 44 & Interferometry &  \citet{Boyajian09} \\
                             & 4911 $\pm$ 75 & IRFM$^{1}$ &  \citet{Schuler09} \\
                             & 4925 $\pm$ 8.7 & Line-depth ratio  (average)&  \citet{Kovtyukh06} \\
                              & 4910 $\pm$ 80 & FeI excitation equilibrium &  \citet{HeMe07} \\
                              & 4925 $\pm$ 84 & Average & Selected results from the PASTEL catalogue \\
\hline
\end{tabular}
\medskip

(1) The effective temperatures adopted by \citet{Schuler09} were calculated by 
\citet{BLG94} using the infrared flux method  for $\gamma$ and $\delta^{1}$ Tau and by \citet{BL98} for $\epsilon$ Tau.
\end{table*}

\subsection{Effective temperatures}


As the only giants of the nearest open cluster, these stars have been analyzed many times. To obtain an idea of the range of temperatures found 
in the literature, previous determinations of the effective temperature (T$_{\rm eff}$) of the four giants were queried at the PASTEL catalogue 
\citep{Soubiran10}. 

A few selected and recent results, together with a ``literature average'', are given in Table \ref{tab:atm} along with the corresponding reference. 
Note that the T$_{\rm eff}$ calculated with the interferometric data by \citet{Boyajian09} is not included in the PASTEL catalogue. 

There is a range of about 300--350 K on the values of T$_{\rm eff}$ determined for each star. As discussed in \citet{Boyajian09}, the values 
determined from recent angular diameters are on the low side of this range. Similarly, using the angular diameter of the Li-rich giant HD148\,193, 
\citet{Baines11} derived an T$_{\rm eff}$ that is on the cooler side of the range of temperatures determined in the literature. Together these results 
might be indicating that the temperature of giants is usually overestimated.

In the following discussion, Na abundances are calculated using two values of temperature for each star: The interferometric temperatures 
of \citet{Boyajian09} and those used by \citet{Schuler09} in their abundance analysis of the same stars. The temperatures of \citet{Schuler09} are close to the average literature values (see Table \ref{tab:atm}) and differ from the interferometric ones by 90 to 150 K.



\subsection{Gravities}

Gravities for the Hyades giants were determined with the equation:

{\small 
\[
log\,(g_{\star}/g_{\odot}) = log\,(M_{\star}/M_{\odot}) + 4\, log\, (T_{\rm eff \star}/T_{\rm  eff  \odot}) - log\, (L_{\star}/L_{\odot})
\]
}

The luminosities by \citet{deBruijne01}, a mass of 2.48 M$_{\odot}$, and 
the usual solar values (T$_{\rm eff \odot}$ = 5777 K and log\,g$_{\odot}$ = 4.44) were adopted. For consistency, gravities were calculated using each of 
the T$_{\rm eff}$ adopted for the analysis. In addition, to illustrate the effect of the gravity in the Na abundance, additional 
values were determined for $\gamma$ Tau assuming masses of 2.0 and 3.0 M$_{\odot}$. 
It can be seen that this change in the mass causes only a minor change on log g, arguing 
that this parameter is well constrained for these stars. 

\subsection{Microturbulence}\label{sec:micro}

Throughout the analysis, a fixed value of microturbulence, $\xi$ = 1.30 km s$^{-1}$ was always adopted. 
This is a typical value found for the giants analyzed in \citet{Sm09b}. This value was checked against empirical relations 
from the literature that calibrate $\xi$ as a function of log g and/or T$_{\rm eff}$. Five calibrations from four references 
were investigated. It is interesting to note that some of these calibrations are given only as a function of log g:

\begin{enumerate}

\item $\xi$ = 2.22 $-$ 0.322 log g \citep[][derived from giants with {[Fe/H]} between $-$1.00 to 0.00]{Gratton96}, 

\item $\xi$ = 1.5 $-$ 0.13 log g \citep[][derived from open cluster giants with {[Fe/H]} between $-$0.37 to +0.24]{CBGT04}, 

\item $\xi$  = 1.645 + (3.854 $\times$ 10$^{-4}$ (T$_{\rm eff}$ $-$ 6387)) + ( $-$0.64 (logg-4.373)) + ($-$3.427 $\times$ 10$^{-4}$ 
(T$_{\rm eff}$ $-$ 6387) (logg $-$ 4.373)) \citep[][derived from solar neighborhood stars with {[Fe/H]} between $-$0.50 to 
+0.50]{AllendePrieto04}, 

\item $\xi$ = 3.40 $-$ 4.41 $\times$ 10$^{-4}$ T$_{\rm eff}$ 
and $\xi$ = 1.84 $-$ 0.202 log g \citep[][derived from bulge, thin and thick disc giants with {[Fe/H]} between $-$1.50 to +0.50]{AlvesBrito10}.

\end{enumerate}

The values obtained from these calibrations for $\gamma$ Tau are given in Table \ref{tab:micro} for the 
parameters using the interferometric and IRFM temperatures. All these calibrations, apart from that of \citet{CBGT04} give 
values that are very close to the one adopted here. What is more important, they show that the variation in 
T$_{\rm eff}$ within the range considered here does not result in a large change in microturbulence. This argues that 
the choice of keeping $\xi$ constant does not introduce systematic effects in the current analysis. 

\begin{table}
\caption{The different sets of atmospheric parameters calculated for the Hyades giants.} \label{tab:par}
\centering
\begin{tabular}{ccccc}
\hline
Star & T$_{\rmn{ eff}}$ & log g & [Fe/H] & Note \\
\hline
 $\gamma$ Tau & 4844 & 2.66 & +0.14 $\pm$ 0.05 & Interferometric T$_{\rmn{ eff}}$ \\
                             & 4844 & 2.57           & +0.13 $\pm$ 0.05 & As above with 2.0 M$_{\odot}$ \\
                              & 4844 & 2.74          &  +0.15 $\pm$ 0.05 & As above with 3.0 M$_{\odot}$ \\
                             & 4965  & 2.70    &  +0.23 $\pm$ 0.05 & IRFM T$_{\rmn{ eff}}$ \\
$\delta^{1}$ Tau & 4826 & 2.69 & +0.18 $\pm$ 0.06 & Interferometric T$_{\rmn{ eff}}$ \\
                             & 4938  &  2.73    & +0.25 $\pm$ 0.06 & IRFM T$_{\rmn{ eff}}$  \\
$\epsilon$ Tau & 4827  & 2.60  & +0.26 $\pm$ 0.08 & Interferometric T$_{\rmn{ eff}}$ \\
                             & 4911 & 2.63     & +0.31 $\pm$ 0.08 & IRFM T$_{\rmn{ eff}}$ \\
\hline
\end{tabular}
\end{table}

\subsection{Metallicity}\label{sec:feh}

To estimate the metallicity of the stars ([Fe/H]), equivalent widths of a set of 15 selected Fe I lines were measured. The line list, atomic 
data, and equivalent widths are given in Table \ref{tab:fe1}. The C$_{\rm 6}$ broadening constants were taken from \citet{CB05}. 
The adopted solar iron abundance is A(Fe) = 7.50 \citep{GS98}. Using 
the interferometric temperatures, the mean metallicity of the Hyades giants is found to be [Fe/H] = +0.19 $\pm$ 0.06. Using the IRFM temperatures, 
this value increases to [Fe/H] = +0.26 $\pm$ 0.04.

\begin{table}
\caption{Microturbulence velocities (in km s$^{-1}$) derived with different calibrations for $\gamma$ Tau, using both the interferometric 
and the IRFM temperature and the corresponding log g.} \label{tab:micro}
\centering
\begin{tabular}{lccc}
\hline
Calibration & Interf. T$_{\rm eff}$  & IRFM T$_{\rm eff}$ \\
\hline
\citet{Gratton96} & 1.37 & 1.35 \\
\citet{AllendePrieto04} & 1.24 & 1.35 \\
\citet{CBGT04} & 1.15 & 1.15 \\
\citet{AlvesBrito10} -- T$_{\rm eff}$ & 1.26 & 1.25 \\
\citet{AlvesBrito10} -- log g & 1.30 & 1.29 \\
\hline
\end{tabular}
\end{table}
\begin{table*}
\caption{Atomic data and equivalent widths of the Fe I lines used to derive the metallicity of the stars.} \label{tab:fe1}
\centering
\begin{tabular}{lccccccc}
\hline
$\lambda$ & $\chi$ & log gf & C$_6$ & Sun & $\gamma$ Tau & $\delta^{1}$ Tau & $\epsilon$ Tau\\ 
 (\AA) & (eV) &  &  & (m\AA) & (m\AA) & (m\AA) & (m\AA)  \\
\hline
5054.642 & 3.64 & $-$2.032 & 4.68E-32 & 40.2 & 80.6 & 80.2 & 84.5 \\
5127.679 & 0.05 & $-$6.005 & 7.38E-33 & 22.3 & 96.6 & 99.5 & 105.6 \\
5223.185 & 3.64 & $-$2.285 & 6.00E-32 & 29.7 & 65.8 & 68.8 & 76.0 \\
5320.035 & 3.64 & $-$2.542 & 8.91E-32 & 19.8 & 57.1 & 57.9 & 59.0 \\
5483.098 & 4.15 & $-$1.481 & 2.95E-31 & 46.5 & 78.8 & 80.8 & 83.1 \\
5522.446 & 4.21 & $-$1.432 & 3.02E-31 & 43.6 & 78.5 & 80.6 & 83.8 \\
5778.453 & 2.59 & $-$3.524 & 4.95E-32 & 22.2 & 73.6 & 74.6 & 81.1 \\
5784.658 & 3.40 & $-$2.626 & 3.57E-31 & 27.2 & 67.8 & 70.7 & 76.2 \\
5814.807 & 4.28 & $-$1.861 & 2.82E-31 & 23.4 & 56.5 & 58.1 & 59.7 \\
6012.209 & 2.22 & $-$3.843 & 3.35E-32 & 24.2 & 78.7 & 78.4 & 82.3 \\
6079.008 & 4.65 & $-$1.055 & 5.13E-31 & 46.2 & 77.8 & 80.9 & 81.2 \\
6187.989 & 3.94 & $-$1.712 & 4.90E-31 & 47.9 & 91.2 & 93.1 & 99.6 \\
6271.278 & 3.33 & $-$2.797 & 2.78E-31 & 23.6 & 68.3 & 74.2 & 79.5 \\
6739.522 & 1.56 & $-$4.942 & 2.10E-32 & 11.8 & 63.6 & 64.8 & 70.3 \\
6837.006 & 4.59 & $-$1.756 & 2.46E-32 & 17.5 & 47.1 & 50.2 & 52.1 \\
\hline
\end{tabular}
\end{table*}
%

\section{Sodium abundances}\label{sec:abun}

\subsection{Line selection and atomic data}

All the 32 strong NaI and NaII lines with wavelengths between 4000 and 8200 \AA\ listed by \citet{SansonettiMartin05} were 
considered as possible features to be used in this analysis. The profiles of these lines were checked both in the UVES\footnote{The 
spectrum is available for download at the ESO website: www.eso.org/observing/dfo/quality/UVES/pipeline/solar\_spec\-trum.html} 
and the \citet{Kurucz05} solar spectra and in the spectrum of $\gamma$ Tau. All lines that were heavily blended or too strong 
for an abundance analysis were discarded (e.g. lines at $\lambda$ 8183 and $\lambda$ 8194 \AA).The atomic 
data of the remaining NaI lines are listed in Table \ref{tab:data}. 
The excitation potential and the log $gf$ of the lines were adopted from the NIST database \citep{NIST10}. The C$_{\rm 6}$ broadening 
constants were adopted from \citet{CB05} and \citet{Bar06}. An assessment of each line is given below:




\begin{enumerate}
\renewcommand{\theenumi}{(\arabic{enumi})}

\item 5\,148.838 \AA: In the solar spectra the line has $\sim$ 14 m\AA. It is slightly blended on the blue wing with a line of 
similar strength. In the $\gamma$ Tau spectrum the two lines are completely blended but their bottoms can be distinguished. Only 
possible to analyze with spectrum synthesis.

\item 5\,682.633 \AA: It is clear in the solar spectra that the blue wing is blended with a weaker line. In the 
spectrum of $\gamma$ Tau the blend can not be recognized, but the line is asymmetric. Equivalent widths 
would be affected by the blend and thus spectrum synthesis should be preferred. In the Sun the line has $\sim$ 100 m\AA. 

\item 5\,688.193 \& 5\,688.205 \AA: These lines are the fine structure components of the same transition. 
The feature is strong in the Sun ($\ga$ 120 m\AA) and stronger in $\gamma$ Tau ($\ga$ 170 m\AA). 
Seems to be clean enough, but it is too strong for an analysis using equivalent widths.

\item 6\,154.226 \AA: On the Sun there is a weak line close to the red wing at $\sim$ 6154.43 \AA. On $\gamma$ Tau the same blend 
is apparent but less distinguishable. There is another weak line at $\sim$ 6154.6 \AA. The line seems to deviate from 
a Gaussian profile towards a Voigt one already in the Sun (where EW $\sim$ 37 m\AA). Analysis 
using equivalent widths should be possible.

\item 6\,160.747 \AA: Both in the Sun and in $\gamma$ Tau, the placement of the continuum is affected by the wing of 
the strong nearby CaI 6162.2 \AA\ line. It has closeby lines to the blue and red sides, but the profile seems clean. Analysis 
using equivalent widths is possible. It has $\sim$ 60 m\AA\ in the Sun.
 


\end{enumerate}

\begin{table}
\caption{Atomic data of the six selected NaI lines.} \label{tab:data}
\centering
\begin{tabular}{lccc}
\hline
 $\lambda$ (\AA) & log gf & $\chi$ (eV) & C$_{6}$ \\
\hline
5\,148.838 & $-$2.044 & 2.102 &  1.01E-30 \\
5\,682.633 & $-$0.706 & 2.102 &  3.38E-30 \\
5\,688.193* & $-$1.406 & 2.104 & 2.03E-30 \\
5\,688.205* & $-$0.452 & 2.104 &  3.38E-30 \\
6\,154.226 & $-$1.547 & 2.102 & 0.90E-31 \\
6\,160.747 & $-$1.246 & 2.104 &  0.30E-31 \\
\hline
\end{tabular}
\medskip

(*) Fine structure components.
\end{table}

\subsection{Equivalent widths}\label{sec:eqw}


Given the above assessment, only the equivalent widths of the lines at 6154 and 6160 \AA\ were 
used to determine Na abundances. However, other lines have been used in the literature. For example, 
\citet{Schuler09} also use the line at 5682 \AA, which 
is clearly blended in the Sun and is asymmetric in $\gamma$ Tau. The equivalent width of such line should 
be regarded as suspicious. Indeed, from the three lines adopted by \citet{Schuler09}, line 5682 \AA\ always results 
in an abundance that is higher by 0.10--0.14 dex than that obtained with the lines at 6154 and 6160 \AA. Without  
this line, the mean [Na/Fe] for the Hyades giants found by \citet{Schuler09} is reduced by 0.05 dex.

Equivalent widths were determined by fitting Gaussian profiles to the observed lines using 
IRAF\footnote{IRAF is distributed by the National Optical Astronomy Observatory, which is operated by the 
Association of Universities for Research in Astronomy, Inc., under cooperative agreement with the National 
Science Foundation.}. For the Sun, lines were measured both in the UVES and in the Kurucz 
spectrum. The values obtained are listed in Table \ref{tab:eqw}.

Model atmospheres were computed using the Linux 
version \citep{Sbordone04,Sbordone05} of the ATLAS9 code originally developed by 
Kurucz \citep[see e.g.][]{Kuruczcd13}. For the calculations, the opacity distribution functions of \citet{ipCK03} 
without overshooting were adopted. These models assume local thermodynamic equilibrium, plane-parallel 
geometry, and hydrostatic equilibrium. Abundances were derived using the WIDTH code, also in its 
Linux version. For the Hyades giants, abundances were calculated only with the equivalent widths measured in the FEROS 
spectra, as they have higher S/N. These Na abundances are listed in Table \ref{tab:naeqw}.

\begin{table*}
\caption{Equivalent widths of three NaI lines measured from the data collected in this work.} \label{tab:eqw}
\centering
\begin{tabular}{lccccccc}
\hline
Line & Sun & Sun  & $\gamma$ Tau & $\delta^{1}$  Tau & $\delta^{1}$  Tau & $\epsilon$ Tau & $\epsilon$ Tau \\
         & Kurucz &  UVES & FEROS & FEROS  & UVES & FEROS & UVES  \\
(\AA) & (m\AA) & (m\AA) & (m\AA) & (m\AA) & (m\AA)  & (m\AA) & (m\AA) \\ 
\hline
6154 & 37.9  & 37.7 & 105.4 & 104.1 & 103.1 & 108.0 & 107.4 \\ 
6160 & 58.1 & 58.6  & 119.6  & 122.8 & 119.8 & 127.0 & 124.8\\
\hline
\end{tabular}
%
\end{table*}
%

%
%
%

%
\begin{table*}
\caption{Sodium abundances calculated with equivalent widths and the different atmospheric parameters of each star.} \label{tab:naeqw}
\centering
\begin{tabular}{lcccccccccc}
\hline
Line  & Sun      &  Sun     & $\gamma$ Tau & $\gamma$ Tau &  $\gamma$ Tau & $\gamma$ Tau & $\delta^{1}$  Tau & $\delta^{1}$  Tau & $\epsilon$ Tau & $\epsilon$ Tau \\
 (\AA)     & Kurucz &  UVES  & Interf.                   & IRFM                  &  2.0 M$_{\odot}$  & 3.0 M$_{\odot}$  & Interf.                   & IRFM                      & Interf.                  & IRFM   \\
\hline
6154  & 6.27 & 6.27 & 6.86 & 6.96 & 6.86 & 6.86 & 6.83 & 6.92 & 6.91 & 6.98 \\ 
6160 & 6.32 & 6.33 & 6.81 & 6.92 &  6.81& 6.81 & 6.86 & 6.96 & 6.95 & 7.02 \\
\hline
Average & 6.29 & 6.30 & 6.84 & 6.94 & 6.84 & 6.84 & 6.85 & 6.94 & 6.93 & 7.00 \\
{[Na/Fe]} & -- & -- & +0.40 & +0.41 & +0.41 & +0.39 & +0.37 & +0.39 & +0.37 & +0.39 \\
\hline
\end{tabular}
%
\end{table*}

\subsection{Spectrum synthesis}

Abundances were also derived using spectrum synthesis and all the Na lines in Table \ref{tab:data}. 
Synthetic spectra were calculated with the codes described in \cite{CB05} and the model atmospheres 
described above. The line list is the one used to compute the spectrum library of \cite{CB05}. As mentioned before, 
the Na line at 5682 \AA\@ is blended in its blue wing. In the line list used here, this blend is due to a Cr I line 
at 5682.495 \AA, with log $gf$ = $-$0.609. It was modeled with the solar abundance recommended by 
\citet{GS98}, A(Cr)$_{\odot}$ = 5.67. The Na abundances obtained with spectrum synthesis are listed in 
Table \ref{tab:naspec}.

\subsection{Uncertainties of the Na abundances}

The main source of error of the abundances are the errors of the atmospheric parameters. 
The uncertainty of the T$_{\rm eff}$ derived by \citet{Boyajian09} is of the order of $\pm$ 50 K. This 
corresponds to an average uncertainty of $\pm$ 0.04 dex in the Na abundance -- A(Na) -- and $\pm$ 0.03 in [Fe/H]. 

As shown for $\gamma$ Tau (Table \ref{tab:par}), gravities are well constrained and no 
significant impact on the Na abundance is expected. A change of $\pm$ 0.05 dex in log g 
results in a change of $\mp$ 0.015 dex in A(Na). The same uncertainty causes an effect of $\mp$ 0.005 in [Fe/H].

The discussion in Section \ref{sec:micro} shows that the value of $\xi$ is well constrained. The values given by the different calibrations have 
an rms of $\pm$ 0.08 km s$^{-1}$. A total change of $\pm$ 0.10 km s$^{-1}$ in $\xi$ results in a 
change of  $\mp$ 0.03 dex in A(Na). {This change causes no significant effect in [Fe/H].}

Considering that there is no significant uncertainty in the Na and Fe abundances of the Sun, the final uncertainty in [Na/Fe] is of $\pm$ 0.04 dex.

\section{Discussion}\label{sec:dis}

The mean solar abundance using equivalent widths is A(Na)$_{\odot}$ = 6.30 while using 
spectrum synthesis it is A(Na)$_{\odot}$ = 6.33. The average [Na/Fe] values for each star 
are given in Tables \ref{tab:naeqw} and \ref{tab:naspec}, respectively for the cases using equivalent widths and 
spectrum synthesis. Although there is some difference in A(Na) between the different temperature scales, 
there is basically no noticeable effect on [Na/Fe].

\begin{table*}
\caption{Sodium abundances with spectrum synthesis and the different atmospheric parameters of each star.} \label{tab:naspec}
\centering
\begin{tabular}{lcccccccc}
\hline
Line & Sun &  Sun  & $\gamma$ Tau  & $\gamma$ Tau & $\delta^{1}$ Tau & $\delta^{1}$ Tau & $\epsilon$ Tau  & $\epsilon$ Tau \\
  (\AA)  & Kurucz & UVES & Interf. & IRFM & Interf. & IRFM & Interf. & IRFM \\
\hline 
5148 & 6.24 & 6.24 & 6.48 & 6.56 & 6.49 & 6.58 & 6.53 & 6.62  \\ 
5682 & 6.36 & 6.36 & 6.87 & 6.97 & 6.82 & 6.92 & 6.94 & 7.00  \\
5688 & 6.42 & 6.42 & 6.88 & 6.98 & 6.83 & 6.94 & 6.98 & 7.05  \\
6154 & 6.30 & 6.30 & 6.75 & 6.86 & 6.73 & 6.83 & 6.80 & 6.88  \\
6160 & 6.32 & 6.32 & 6.81 & 6.91 & 6.85 & 6.96 & 6.93 & 7.00  \\
\hline
Average & 6.33 & 6.33 & 6.76 & 6.86 & 6.74 & 6.85 & 6.84 & 6.91 \\
{[Na/Fe]} & -- & -- & +0.29 & +0.30 & +0.23 & +0.27 & +0.25 & +0.27 \\
\hline
\end{tabular}
\end{table*}

However, at first glance the average values determined with equivalent widths seem to be about $\sim$0.12 dex 
higher than the values determined using spectrum synthesis. The difference is apparent in the giants 
but not in the Sun. 

As discussed below, this is not directly related to the use equivalent widths or spectrum synthesis to derive the abundances, but to the different choice of 
lines used in each case.  There is however, a systematic difference of about $\sim$ 0.10 dex between the abundances 
derived with the $\lambda$ 6154 line. This likely comes from an uncertainty in the equivalent width of this line. An error on the 
equivalent width of $\pm$ 6m\AA\ can produce a change of $\pm$ 0.10 on the abundance.

There are three effects contributing to the difference. One, is the Na abundance given by the $\lambda$ 5148 
line. Excluding it from the average, the [Na/Fe] values increase by $\sim$ 0.05 dex. It is not clear why 
this line results systematically in smaller values. Nevertheless, it was decided to consider its abundance as 
suspicious and to exclude it from further discussion. The final average LTE [Na/Fe] values, without line 
$\lambda$ 5148, are given in Table \ref{tab:nafe}. A second effect discussed in 
the next subsection are the NLTE corrections.
 
The last effect seems related to differences between the WIDTH code, used to calculate abundances from 
equivalent widths, and the PFANT code, used to compute the synthetic spectrum.

Following a suggestion by the referee, a series of synthetic spectra with the parameters of $\gamma$ Tau at the region 
around the line $\lambda$ 6160 were calculated with different Na abundances. The equivalent 
widths were measured and the values used to recompute the abundances. The 
resulting Na abundances given by WIDTH are 0.03 to 0.04 dex higher than the values used to compute the 
synthetic spectra.

At this point, it is not possible to say whether this is caused by some numerical 
effect, by some difference in other input data (e.g. partition functions, opacities...), or to some difference 
in the physics, like the treatment of broadening, for example. This should, of course, be further investigated. Nevertheless, 
as it is based in the analysis of more features and likely based in a more reliable way to deal with the effects of broadening, 
the abundances derived with spectrum synthesis are preferred here.

\subsection{NLTE abundances}

As mentioned in the introduction, the Na lines are affected by NLTE effects \citep[see][and references therein]{Lind11b}. 
This has been suggested as a likely reason behind the large Na overabundances found in giants of 
open clusters \citep[see e.g.][]{R06,Ses07}.

To correct the abundances calculated here, the NLTE calculations of \citet{Lind11b} were adopted.  Corrections were 
interpolated among the grid calculated in that work with an IDL routine kindly made available to the author by 
Karin Lind (2011, private communication). NLTE corrections were derived in a line-by-line basis, giving as input the 
atmospheric parameters and the LTE abundances\footnote{Alternatively, the interpolation code can 
accept as input the equivalent width of the line and return both the LTE and the NLTE abundances. The first approach was preferred 
here because the LTE abundances calculated in \citet{Lind11b} for a given set of atmospheric parameters are 
slightly different from the ones derived in this work. The difference is caused by different choices made 
in the treatment of line broadening.}. 

Individual line corrections for all stars (giants and Sun) range from $-$0.03 (for line 5148 \AA) up to $-$0.15 (for line 6160 \AA). For the Sun, 
the average corrections are of $-$0.08 and $-$0.09 dex, respectively with the abundances using equivalent widths and spectrum 
synthesis. Thus, in NLTE, the solar Na abundances derived in this work are A(Na)$_{\odot}$ = 6.22 with equivalent widths 
and A(Na)$_{\odot}$ = 6.24 with spectrum synthesis. For the giants, the different selection of lines results on average corrections of $-$0.14 when using 
equivalent widths and around $-$0.11 when using spectrum synthesis. 

This difference in the average the NLTE corrections is another responsible for the difference among the LTE [Na/Fe] values given by 
equivalent widths and spectrum synthesis. With respect to the Sun, the correction when using equivalent widths (only lines 6154 and 6160 \AA) 
is of $\sim$ $-$0.06 dex. When using spectrum synthesis (4 lines), the correction is of $-$0.01 or $-$0.02 dex. This helps to explain why the 
different NLTE abundances in Table \ref{tab:nanlte} are in better agreement than the LTE values in Table \ref{tab:nafe}.

The average absolute NLTE Na abundance of the Hyades giants, using the interferometric temperatures, 
is found to be [Na/Fe] = +0.30\footnote{Using equivalent widths the value is [Na/Fe] = +0.33.}. 
With the IRFM adopted by \citet{Schuler09} the value found here is [Na/Fe] = +0.31\footnote{Using 
equivalent widths the value is [Na/Fe] = +0.34.}. 

\begin{table}
\caption{Average [Na/Fe], in LTE, for each star and for each of the different analysis and parameters (excluding line $\lambda$ 5148 for the synthesis values).} \label{tab:nafe}
\centering
\begin{tabular}{lccc}
\hline
Analysis  & $\gamma$ Tau & $\delta^{1}$ Tau & $\epsilon$ Tau \\
\hline
EqW \& Interf. & +0.40 & +0.37 & +0.37 \\
EqW \& IRFM & +0.41 & +0.39 & +0.39 \\
Synthesis \& Interf & +0.34 & +0.28 & +0.30 \\
Synthesis \& IRFM & +0.35 & +0.31 & +0.32 \\
\hline
\end{tabular}
\end{table}

Using equivalent widths, \citet{Schuler09} obtained [Na/Fe] = +0.48 in LTE. The question then is why the 
results are different. First, as pointed out in Section \ref{sec:eqw}, \citet{Schuler09} used the equivalent 
width of line $\lambda$ 5682 to determine the Na abundance. This line, however, is blended and removing it 
from the average reduces [Na/Fe] by 0.05 dex. Second are the NLTE corrections for lines $\lambda$ 6154 and 
6160, causing another reduction by 0.05/0.06 dex. Last, \citet{Schuler09} adopted [Fe/H] = +0.13 for the Hyades 
giants. The atmospheric parameters used in \citet{Schuler09} were determined in a previous paper, \citet{Schu06}, 
were FeI lines were also measured. Using these lines, \citet{Schu06} found an average of [Fe/H] = +0.16, although 
recalculating it with the Kurucz model atmospheres and codes used here a value of [Fe/H] = +0.19 is found. Taking into 
account this difference the final [Na/Fe] in NLTE found by \citet{Schuler09} should be [Na/Fe] = +0.34 
\citep[in the][metallicity scale]{Schu06} or [Na/Fe] = +0.31 (in the metallicity scale recalculated here). These values 
are in perfect agreement with the ones derived in this work.

\begin{table}
\caption{Average [Na/Fe] in NLTE.} \label{tab:nanlte}
\centering
\begin{tabular}{lccc}
\hline
 Analysis & $\gamma$ Tau & $\delta^{1}$ Tau & $\epsilon$ Tau \\
\hline
EqW \& Interf. & +0.34 & +0.33 & +0.32 \\
EqW \& IRFM & +0.35 & +0.33 & +0.34 \\
Synthesis \& Interf & +0.33 & +0.27 & +0.29 \\
Synthesis \& IRFM & +0.33 & +0.29 & +0.31 \\
\hline
\end{tabular}
\end{table}

\subsection{Comparison with evolutionary models}

An interesting question to look at now is whether the Na overabundances of the Hyades giants 
can be explained by evolutionary models. In standard models, no modification of the Na abundance is 
expected after the first dredge-up for stars below $\sim$ 2.0 M$_{\odot}$ \citep{Mow99,CharbonnelLagarde10}. 
For stars of higher mass, an increase of up to +0.20 dex in the Na abundance is expected.

When mixing induced by rotation is included in the models (transport of chemicals and angular momentum by shear turbulence and 
meridional circulation), larger overabundances are produced. These effects can also create a dispersion in the 
Na abundance among otherwise similar stars if they had different initial rotation velocities in the zero age main sequence 
\citep{CharbonnelLagarde10}. This happens because rotation affects the internal abundance profile of the elements 
involved in H-burning. In this way, the Na-rich region in rotating stars begins further out from the core, and 
more Na-rich material can be dredged-up to the surface.

\begin{figure}
\centering
\includegraphics[width=65mm]{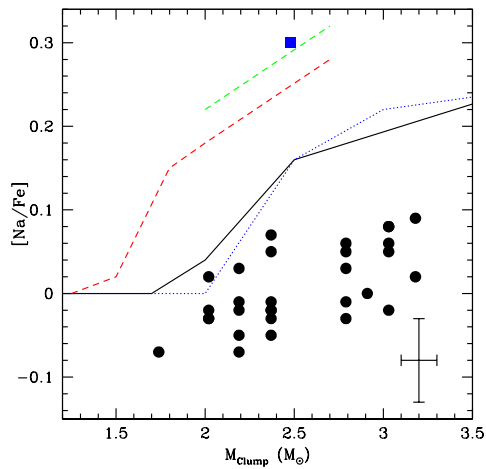}
\caption{Sodium abundances, [Na/Fe], as function of the stellar mass at the clump. The circles indicate the open cluster giants analyzed in \citet{Sm09b}, the 
typical error bar of that work is shown in the lower right corner. The blue square corresponds to the recommended [Na/Fe] of the Hyades 
derived in this work. The lines represent the predicted [Na/Fe] as a function of initial stellar mass given by the models of 
\citet{CharbonnelLagarde10} for the standard case (solid line), for a model 
including thermohaline mixing only (blue dotted line), and models with thermohaline mixing and rotation-induced mixing with initial velocities in the ZAMS 
of 250 and 300 km s$^{-1}$ (lower red dashed line and upper green dashed line, respectively).}
\label{fig:nasm09}
\end{figure}

In Fig.\ \ref{fig:nasm09} the Na abundance of the Hyades is shown in comparison with the models 
calculated by \citet{CharbonnelLagarde10}. Also shown in the figure are the Na abundances of 31 giants 
of 10 open clusters derived in \citet{Sm09b}. These clusters have turn-off masses between 1.7 and 3.1 
M$_{\odot}$. The recommended Na abundance of the Hyades ([Na/Fe] = +0.30) is in the upper part of the range expected by 
the models. Therefore, and contrary to the 
conclusion of \citet{Schuler09}, this comparison shows that the Na abundance in the Hyades 
can be explained by modern hydrodynamical models that include the effects of rotation. There 
is no need for an extra unknown mixing process.

Although it is believed here that the absolute Na abundance of the Hyades was derived, 
this claim can not be extended to most of the results in the literature. One way to avoid such systematics 
and test whether the Na overabundances in giants conform with the prediction of the models, is to 
conduct a large homogeneous analysis of a sample including only giants but with different masses. In this way one can look for abundance 
trends with mass and test if they agree with the expectations of models. 

This is exemplified by the giants analyzed in \citet{Sm09b}. As seen in Fig.\ \ref{fig:nasm09} and discussed also in \citet{Sm09b} 
and \citet{CharbonnelLagarde10}, there is an off-set of about 0.10 dex between observations and models. However, there 
is an agreement in the increasing trend with mass. This suggests that we are indeed observing the effects of mixing in these stars, in spite of a 
possible systematic effect in the abundance scale. 

The Na abundance of the Hyades derived here and the trend of increasing Na with increasing mass found in \citet{Sm09b} 
argue that there is no strong evidence for overabundances above those expected by the models, for stars above 2 M$_{\odot}$. 

At this point, it is interesting to mention the results of \citet{PRZ04} for the cluster IC 4651 and of \citet{R06} for M67. 
\citet{PRZ04} found a systematic difference at the level of 0.20 dex between the Na abundances of dwarfs and giants in IC 4651. 
A similar difference between dwarfs and giants was, however, not detected in stars of M67 by \citet{R06}. Although these results 
could be regarded as contradictory at first sight, they are not. Clump giants in M67 have $\sim$ 1.3 M$_{\odot}$ while clump giants in 
IC 4651 have $\sim$ 1.8 M$_{\odot}$. As can be seen in Fig.\ \ref{fig:nasm09}, according to the models of \citet{CharbonnelLagarde10} 
a star of 1.3 M$_{\odot}$ is never expected to enrich itself in Na after the first dredge-up while stars of 1.8 M$_{\odot}$ could be 
enriched by $\sim$ +0.15 dex. Although caution is needed in the comparison among dwarfs and giants, these results seem to support the 
idea that models with rotation can properly explain the behaviour of Na also in giants below 2 M$_{\odot}$.

\subsection{Cluster versus field giants}

It is sometimes noticed that the overabundance of Na seen in giants of open clusters is not apparent in field 
stars \citep{ipFriel06}. This comparison is, however, usually made between giants in clusters and field 
dwarfs \citep[see e.g.][]{Jacobson11}. As also discussed by these authors, such an offset might be real and caused by mixing. 
Based on the discussion of the previous Section, this 
is a conclusion supported here. \citet{Jacobson11} noticed indeed that their cluster 
abundances agree well with the Na abundances of field clump giants determined by \citet{Mi06}.

Comparisons among dwarfs and giants need to be careful not just because of mixing, but also because systematic effects can cause 
biases that might be mistaken by real differences. As an example, according to \citet{Melendez08} and \citet{AlvesBrito10} this seems to be 
the case behind previous claims of abundance differences among thick disc and bulge stars.



A mismatch of Na abundances between stars in open clusters and in the field lead \citet{DeSilva09} to 
suggest that the dissolution of open clusters might not be the main contributor of stars for the Galactic disc. 
These authors compiled Na abundances in cluster giants from the literature, normalized them to a common solar 
scale, and did a comparison with Na abundances in field dwarfs from \citet{SG05}, field clump giants from \citet{Mi06}, 
and bulge giants from \citet*{FMR07}. Agreement was found between the cluster giants and the field giants of \citet{Mi06}, 
but an offset exists with respect to the dwarfs of \citet{SG05} and the giants of \citet{FMR07}. 


Because of mixing, an offset between Na in giants and dwarfs might be expected, with the caveat that in samples of 
field stars one is adding together stars with different masses and metallicities, properties that affect the mixing of 
Na during the first dredge-up. In other words, depending on the mass range of the giants 
an offset between the Na abundances between giants and dwarfs and among giants themselves might be expected or not. A robust way to attempt a comparison such as the 
one done by \citet{DeSilva09} would be using dwarfs in the field and dwarfs in clusters, where mixing is not able to 
affect the Na abundances.




\section{Summary}\label{sec:con}

Sodium abundances of three Hyades giants have been redetermined and an average value 
of [Na/Fe] = +0.30 in NLTE was found. This Na abundance was derived using the absolute 
T$_{\rm eff}$ of the stars determined with the interferometric angular diameter measurements of 
\citet{Boyajian09}. 

This Na abundance agrees well with the ones predicted by the hydrodynamical models of \citet{CharbonnelLagarde10} for a star of 2.48 M$_{\odot}$, 
after the first dredge-up, and taking into account rotation-induced mixing. This contradicts 
the conclusion of \citet{Schuler09} that the Na overabundances of the Hyades could not be 
explained by any known mixing mechanism. The Na abundance of the Hyades giants are on the upper-limit of the range 
predicted by the models, implying that the stars had a rather high initial rotation.

Absolute abundance values have always to be considered with care. Nevertheless, as a fundamental temperature was used, the Na abundance 
derived here should be quite accurate. In general, relative 
comparisons should be more robust. In this sense, the increasing trend of the Na abundance with increasing 
mass found in the giants analyzed by \citet{Sm09b} is the same as the one expected by 
evolutionary models. Agreement with models is also seen in the results of \citet{R06}, that 
did not find a difference in Na among dwarfs and giants of M67 and in the results of 
\citet{PRZ04}, that did find a difference in Na among dwarfs and giants of IC 4651. 
In addition, it should be noticed that when comparing field giants with cluster giants, similar Na overabundances 
are apparent \citep{Mi06}. 

All these pieces of evidence seem to point to the conclusion that, so far, 
there seems to be no strong evidence for Na overabundances in giants of open 
clusters beyond those that can be well explained by the effects of evolutionary mixing, in stars more massive 
than $\sim$ 2.0 M$_{\odot}$. A consistent and homogeneous reanalysis of Na abundances in a large sample of giants 
is still necessary to confirm (or refute) this conclusion. 

\section*{Acknowledgments}

I thank the anonymous referee for the valuable suggestions and comments. I am also grateful 
to Luca Pasquini for making available the FEROS spectra of the Hyades giants and to Karin Lind 
for making available the code to interpolate among the grid of NLTE corrections. The 
research leading to these results has received funding from the European Community's Seventh Framework 
Programme (FP7/2007-2013) under grant agreement No 229517. This research 
has made use of the WEBDA database, operated at the Institute for Astronomy of the University of Vienna, 
of the Simbad database operated at CDS, Strasbourg, France, and of NASA's Astrophysics Data System.

\bibliographystyle{mn2e}
\bibliography{../../Tese/rsmiljanic}

\begin{thebibliography}{}

\bibitem[\protect\citeauthoryear{{Allende Prieto}, {Barklem}, {Lambert} \&
  {Cunha}}{{Allende Prieto} et~al.}{2004}]{AllendePrieto04}
{Allende Prieto} C.,  {Barklem} P.~S.,  {Lambert} D.~L.,    {Cunha} K.,  2004,
  \aap, 420, 183

\bibitem[\protect\citeauthoryear{{Alves-Brito}, {Mel{\'e}ndez}, {Asplund},
  {Ram{\'{\i}}rez} \& {Yong}}{{Alves-Brito} et~al.}{2010}]{AlvesBrito10}
{Alves-Brito} A.,  {Mel{\'e}ndez} J.,  {Asplund} M.,  {Ram{\'{\i}}rez} I.,
  {Yong} D.,  2010, \aap, 513, A35+

\bibitem[\protect\citeauthoryear{{Andrievsky}, {Spite}, {Korotin}, {Spite},
  {Bonifacio}, {Cayrel}, {Hill} \& {Fran\c cois}}{{Andrievsky}
  et~al.}{2007}]{And07b}
{Andrievsky} S.~M.,  {Spite} M.,  {Korotin} S.~A.,  {Spite} F.,  {Bonifacio}
  P.,  {Cayrel} R.,  {Hill} V.,    {Fran\c cois} P.,  2007, \aap, 464, 1081

\bibitem[\protect\citeauthoryear{{Angelou}, {Church}, {Stancliffe}, {Lattanzio}
  \& {Smith}}{{Angelou} et~al.}{2011}]{Angelou11}
{Angelou} G.~C.,  {Church} R.~P.,  {Stancliffe} R.~J.,  {Lattanzio} J.~C.,
  {Smith} G.~H.,  2011, \apj, 728, 79

\bibitem[\protect\citeauthoryear{{Baines}, {McAlister}, {ten Brummelaar},
  {Turner}, {Sturmann}, {Sturmann}, {Goldfinger}, {Farrington} \&
  {Ridgway}}{{Baines} et~al.}{2011}]{Baines11}
{Baines} E.~K.,  {McAlister} H.~A.,  {ten Brummelaar} T.~A.,  {Turner} N.~H.,
  {Sturmann} J.,  {Sturmann} L.,  {Goldfinger} P.~J.,  {Farrington} C.~D.,
  {Ridgway} S.~T.,  2011, \apj, 731, 132

\bibitem[\protect\citeauthoryear{{Ballester}, {Modigliani}, {Boitquin},
  {Cristiani}, {Hanuschik}, {Kaufer} \& {Wolf}}{{Ballester}
  et~al.}{2000}]{uvespipeline}
{Ballester} P.,  {Modigliani} A.,  {Boitquin} O.,  {Cristiani} S.,  {Hanuschik}
  R.,  {Kaufer} A.,    {Wolf} S.,  2000, The Messenger, 101, 31

\bibitem[\protect\citeauthoryear{{Barbuy}, {Zoccali}, {Ortolani}, {Momany},
  {Minniti}, {Hill}, {Renzini}, {Rich}, {Bica}, {Pasquini} \& {Yadav}}{{Barbuy}
  et~al.}{2006}]{Bar06}
{Barbuy} B.,  {Zoccali} M.,  {Ortolani} S.,  {Momany} Y.,  {Minniti} D.,
  {Hill} V.,  {Renzini} A.,  {Rich} R.~M.,  {Bica} E.,  {Pasquini} L.,
  {Yadav} R.~K.~S.,  2006, \aap, 449, 349

\bibitem[\protect\citeauthoryear{{Baumueller}, {Butler} \&
  {Gehren}}{{Baumueller} et~al.}{1998}]{BBG98}
{Baumueller} D.,  {Butler} K.,    {Gehren} T.,  1998, \aap, 338, 637

\bibitem[\protect\citeauthoryear{{Blackwell} \& {Lynas-Gray}}{{Blackwell} \&
  {Lynas-Gray}}{1994}]{BLG94}
{Blackwell} D.~E.,  {Lynas-Gray} A.~E.,  1994, \aap, 282, 899

\bibitem[\protect\citeauthoryear{{Blackwell} \& {Lynas-Gray}}{{Blackwell} \&
  {Lynas-Gray}}{1998}]{BL98}
{Blackwell} D.~E.,  {Lynas-Gray} A.~E.,  1998, \aaps, 129, 505

\bibitem[\protect\citeauthoryear{{Boyajian}, {McAlister}, {Cantrell}, {Gies},
  {Brummelaar}, {Farrington}, {Goldfinger}, {Sturmann}, {Sturmann}, {Turner} \&
  {Ridgway}}{{Boyajian} et~al.}{2009}]{Boyajian09}
{Boyajian} T.~S.,  {McAlister} H.~A.,  {Cantrell} J.~R.,  {Gies} D.~R.,
  {Brummelaar} T.~A.~t.,  {Farrington} C.,  {Goldfinger} P.~J.,  {Sturmann} L.,
   {Sturmann} J.,  {Turner} N.~H.,    {Ridgway} S.,  2009, \apj, 691, 1243

\bibitem[\protect\citeauthoryear{{Bragaglia}, {Carretta}, {Gratton}, {Tosi},
  {Bonanno}, {Bruno}, {Cal{\`i}}, {Claudi}, {Cosentino}, {Desidera},
  {Farisato}, {Rebeschini} \& {Scuderi}}{{Bragaglia} et~al.}{2001}]{Brag01}
{Bragaglia} A.,  {Carretta} E.,  {Gratton} R.~G.,  {Tosi} M.,  {Bonanno} G.,
  {Bruno} P.,  {Cal{\`i}} A.,  {Claudi} R.,  {Cosentino} R.,  {Desidera} S.,
  {Farisato} G.,  {Rebeschini} M.,    {Scuderi} S.,  2001, \aj, 121, 327

\bibitem[\protect\citeauthoryear{{Canto Martins}, {L{\`e}bre}, {Palacios}, {de
  Laverny}, {Richard}, {Melo}, {Do Nascimento} Jr. \& {de Medeiros}}{{Canto
  Martins} et~al.}{2011}]{CantoMartins11}
{Canto Martins} B.~L.,  {L{\`e}bre} A.,  {Palacios} A.,  {de Laverny} P.,
  {Richard} O.,  {Melo} C.~H.~F.,  {Do Nascimento} Jr. J.~D.,    {de Medeiros}
  J.~R.,  2011, \aap, 527, A94+

\bibitem[\protect\citeauthoryear{{Carretta}, {Bragaglia}, {Gratton} \&
  {Tosi}}{{Carretta} et~al.}{2004}]{CBGT04}
{Carretta} E.,  {Bragaglia} A.,  {Gratton} R.~G.,    {Tosi} M.,  2004, \aap,
  422, 951

\bibitem[\protect\citeauthoryear{{Castelli} \& {Kurucz}}{{Castelli} \&
  {Kurucz}}{2003}]{ipCK03}
{Castelli} F.,  {Kurucz} R.~L.,  2003, in {Piskunov} N.,  {Weiss} W.~W.,
  {Gray} D.~F.,  eds, Proceedings of the IAU Symposium 210: Modelling of
  Stellar Atmospheres {New Grids of ATLAS9 Model Atmospheres Modelling of
  Stellar Atmospheres}.
ASP, San Francisco, p.~A20

\bibitem[\protect\citeauthoryear{{Charbonnel}, {Brown} \&
  {Wallerstein}}{{Charbonnel} et~al.}{1998}]{CBW98}
{Charbonnel} C.,  {Brown} J.~A.,    {Wallerstein} G.,  1998, \aap, 332, 204

\bibitem[\protect\citeauthoryear{{Charbonnel} \& {Lagarde}}{{Charbonnel} \&
  {Lagarde}}{2010}]{CharbonnelLagarde10}
{Charbonnel} C.,  {Lagarde} N.,  2010, \aap, 522, A10+

\bibitem[\protect\citeauthoryear{{Charbonnel} \& {Talon}}{{Charbonnel} \&
  {Talon}}{2008}]{ipCT08}
{Charbonnel} C.,  {Talon} S.,  2008, in {Deng} L.,  {Chan} K.~L.,  eds,
  Proceedings of the IAU Symposium 252 {Deep inside low-mass stars}.
pp 163--174

\bibitem[\protect\citeauthoryear{{Coelho}, {Barbuy}, {Mel\'endez}, {Schiavon}
  \& {Castilho}}{{Coelho} et~al.}{2005}]{CB05}
{Coelho} P.,  {Barbuy} B.,  {Mel\'endez} J.,  {Schiavon} R.~P.,    {Castilho}
  B.~V.,  2005, \aap, 443, 735

\bibitem[\protect\citeauthoryear{{de Bruijne}, {Hoogerwerf} \& {de Zeeuw}}{{de
  Bruijne} et~al.}{2001}]{deBruijne01}
{de Bruijne} J.~H.~J.,  {Hoogerwerf} R.,    {de Zeeuw} P.~T.,  2001, \aap, 367,
  111

\bibitem[\protect\citeauthoryear{{de Silva}, {Gibson}, {Lattanzio} \&
  {Asplund}}{{de Silva} et~al.}{2009}]{DeSilva09}
{de Silva} G.~M.,  {Gibson} B.~K.,  {Lattanzio} J.,    {Asplund} M.,  2009,
  \aap, 500, L25

\bibitem[\protect\citeauthoryear{{Dekker}, {D'Odorico}, {Kaufer}, {Delabre} \&
  {Kotzlowski}}{{Dekker} et~al.}{2000}]{ipDe00}
{Dekker} H.,  {D'Odorico} S.,  {Kaufer} A.,  {Delabre} B.,    {Kotzlowski} H.,
  2000, in {Iye} M.,  {Moorwood} A.~F.,  eds, Proc. SPIE, Optical and IR
  Telescope Instrumentation and Detectors Vol.~4008, {Design, construction, and
  performance of UVES, the echelle spectrograph for the UT2 Kueyen Telescope at
  the ESO Paranal Observatory}.
SPIE, pp 534--545

\bibitem[\protect\citeauthoryear{{Denisenkov} \& {Denisenkova}}{{Denisenkov} \&
  {Denisenkova}}{1990}]{DD90}
{Denisenkov} P.~A.,  {Denisenkova} S.~N.,  1990, Soviet Astronomy Letters, 16,
  275

\bibitem[\protect\citeauthoryear{{Denissenkov} \& {Pinsonneault}}{{Denissenkov}
  \& {Pinsonneault}}{2008}]{DePi08}
{Denissenkov} P.~A.,  {Pinsonneault} M.,  2008, \apj, 684, 626

\bibitem[\protect\citeauthoryear{{Denissenkov}, {Pinsonneault} \&
  {MacGregor}}{{Denissenkov} et~al.}{2009}]{Den09}
{Denissenkov} P.~A.,  {Pinsonneault} M.,    {MacGregor} K.~B.,  2009, \apj,
  696, 1823

\bibitem[\protect\citeauthoryear{{Friel}}{{Friel}}{2006}]{ipFriel06}
{Friel} E.~D.,  2006, in {Randich} S.,  {Pasquini} L.,  eds, Chemical
  Abundances and Mixing in Stars in the Milky Way and its Satellites, ESO
  Astrophysics Symposia {Metallicities and {$\alpha$}-Abundancesin Open
  Clusters}.
p.~3

\bibitem[\protect\citeauthoryear{{Friel}, {Jacobson}, {Barrett}, {Fullton},
  {Balachandran} \& {Pilachowski}}{{Friel} et~al.}{2003}]{Fri03}
{Friel} E.~D.,  {Jacobson} H.~R.,  {Barrett} E.,  {Fullton} L.,  {Balachandran}
  S.~C.,    {Pilachowski} C.~A.,  2003, \aj, 126, 2372

\bibitem[\protect\citeauthoryear{{Friel}, {Jacobson} \& {Pilachowski}}{{Friel}
  et~al.}{2010}]{Friel10}
{Friel} E.~D.,  {Jacobson} H.~R.,    {Pilachowski} C.~A.,  2010, \aj, 139, 1942

\bibitem[\protect\citeauthoryear{{Fulbright}, {McWilliam} \&
  {Rich}}{{Fulbright} et~al.}{2007}]{FMR07}
{Fulbright} J.~P.,  {McWilliam} A.,    {Rich} R.~M.,  2007, \apj, 661, 1152

\bibitem[\protect\citeauthoryear{{Geisler}, {Smith}, {Wallerstein}, {Gonzalez}
  \& {Charbonnel}}{{Geisler} et~al.}{2005}]{Gei05}
{Geisler} D.,  {Smith} V.~V.,  {Wallerstein} G.,  {Gonzalez} G.,
  {Charbonnel} C.,  2005, \aj, 129, 1428

\bibitem[\protect\citeauthoryear{{Gilroy}}{{Gilroy}}{1989}]{G89}
{Gilroy} K.~K.,  1989, \apj, 347, 835

\bibitem[\protect\citeauthoryear{{Gilroy} \& {Brown}}{{Gilroy} \&
  {Brown}}{1991}]{GB91}
{Gilroy} K.~K.,  {Brown} J.~A.,  1991, \apj, 371, 578

\bibitem[\protect\citeauthoryear{{Girardi}, {Bertelli}, {Bressan}, {Chiosi},
  {Groenewegen}, {Marigo}, {Salasnich} \& {Weiss}}{{Girardi}
  et~al.}{2002}]{Gir02}
{Girardi} L.,  {Bertelli} G.,  {Bressan} A.,  {Chiosi} C.,  {Groenewegen}
  M.~A.~T.,  {Marigo} P.,  {Salasnich} B.,    {Weiss} A.,  2002, \aap, 391, 195

\bibitem[\protect\citeauthoryear{{Gratton}, {Carretta} \& {Castelli}}{{Gratton}
  et~al.}{1996}]{Gratton96}
{Gratton} R.~G.,  {Carretta} E.,    {Castelli} F.,  1996, \aap, 314, 191

\bibitem[\protect\citeauthoryear{{Gratton}, {Carretta}, {Eriksson} \&
  {Gustafsson}}{{Gratton} et~al.}{1999}]{GCEG99}
{Gratton} R.~G.,  {Carretta} E.,  {Eriksson} K.,    {Gustafsson} B.,  1999,
  \aap, 350, 955

\bibitem[\protect\citeauthoryear{{Gratton}, {Sneden}, {Carretta} \&
  {Bragaglia}}{{Gratton} et~al.}{2000}]{Gr00}
{Gratton} R.~G.,  {Sneden} C.,  {Carretta} E.,    {Bragaglia} A.,  2000, \aap,
  354, 169

\bibitem[\protect\citeauthoryear{{Grevesse} \& {Sauval}}{{Grevesse} \&
  {Sauval}}{1998}]{GS98}
{Grevesse} N.,  {Sauval} A.~J.,  1998, Space Science Reviews, 85, 161

\bibitem[\protect\citeauthoryear{{Griffin}, {Griffin}, {Gunn} \&
  {Zimmerman}}{{Griffin} et~al.}{1988}]{Griffin88}
{Griffin} R.~F.,  {Griffin} R.~E.~M.,  {Gunn} J.~E.,    {Zimmerman} B.~A.,
  1988, \aj, 96, 172

\bibitem[\protect\citeauthoryear{{Hamdani}, {North}, {Mowlavi}, {Raboud} \&
  {Mermilliod}}{{Hamdani} et~al.}{2000}]{Ham00}
{Hamdani} S.,  {North} P.,  {Mowlavi} N.,  {Raboud} D.,    {Mermilliod} J.-C.,
  2000, \aap, 360, 509

\bibitem[\protect\citeauthoryear{{Hekker} \& {Mel{\'e}ndez}}{{Hekker} \&
  {Mel{\'e}ndez}}{2007}]{HeMe07}
{Hekker} S.,  {Mel{\'e}ndez} J.,  2007, \aap, 475, 1003

\bibitem[\protect\citeauthoryear{{Iben}}{{Iben}}{1967}]{I67}
{Iben} I.~J.,  1967, \apj, 147, 624

\bibitem[\protect\citeauthoryear{{Jacobson}, {Friel} \&
  {Pilachowski}}{{Jacobson} et~al.}{2007}]{Jac07}
{Jacobson} H.~R.,  {Friel} E.~D.,    {Pilachowski} C.~A.,  2007, \aj, 134, 1216

\bibitem[\protect\citeauthoryear{{Jacobson}, {Friel} \&
  {Pilachowski}}{{Jacobson} et~al.}{2008}]{Jac08}
{Jacobson} H.~R.,  {Friel} E.~D.,    {Pilachowski} C.~A.,  2008, \aj, 135, 2341

\bibitem[\protect\citeauthoryear{{Jacobson}, {Friel} \&
  {Pilachowski}}{{Jacobson} et~al.}{2011}]{Jacobson11}
{Jacobson} H.~R.,  {Friel} E.~D.,    {Pilachowski} C.~A.,  2011, \aj, 141, 58

\bibitem[\protect\citeauthoryear{{Kaufer}, {Stahl}, {Tubbesing}, {Norregaard},
  {Avila}, {Francois}, {Pasquini} \& {Pizzella}}{{Kaufer} et~al.}{1999}]{Kau99}
{Kaufer} A.,  {Stahl} O.,  {Tubbesing} S.,  {Norregaard} P.,  {Avila} G.,
  {Francois} P.,  {Pasquini} L.,    {Pizzella} A.,  1999, The Messenger, 95, 8

\bibitem[\protect\citeauthoryear{{Kovtyukh}, {Soubiran}, {Bienaym{\'e}},
  {Mishenina} \& {Belik}}{{Kovtyukh} et~al.}{2006}]{Kovtyukh06}
{Kovtyukh} V.~V.,  {Soubiran} C.,  {Bienaym{\'e}} O.,  {Mishenina} T.~V.,
  {Belik} S.~I.,  2006, \mnras, 371, 879

\bibitem[\protect\citeauthoryear{{Kurucz}}{{Kurucz}}{1993}]{Kuruczcd13}
{Kurucz} R.,  1993, ATLAS9 Stellar Atmosphere Programs and 2 km/s grid. CD-ROM
  No.~13.~ Cambridge, Mass.: Smithsonian Astrophysical Observatory.

\bibitem[\protect\citeauthoryear{{Kurucz}}{{Kurucz}}{2005}]{Kurucz05}
{Kurucz} R.~L.,  2005, Memorie della Societa Astronomica Italiana Supplementi,
  8, 189

\bibitem[\protect\citeauthoryear{{L{\`e}bre}, {de Laverny}, {De Medeiros},
  {Charbonnel} \& {da Silva}}{{L{\`e}bre} et~al.}{1999}]{Leb99}
{L{\`e}bre} A.,  {de Laverny} P.,  {De Medeiros} J.~R.,  {Charbonnel} C.,
  {da Silva} L.,  1999, \aap, 345, 936

\bibitem[\protect\citeauthoryear{{Lind}, {Asplund}, {Barklem} \&
  {Belyaev}}{{Lind} et~al.}{2011}]{Lind11b}
{Lind} K.,  {Asplund} M.,  {Barklem} P.~S.,    {Belyaev} A.~K.,  2011, \aap,
  528, A103+

\bibitem[\protect\citeauthoryear{{Mashonkina}, {Shimanski{\u i}} \&
  {Sakhibullin}}{{Mashonkina} et~al.}{2000}]{MSS00}
{Mashonkina} L.~I.,  {Shimanski{\u i}} V.~V.,    {Sakhibullin} N.~A.,  2000,
  Astronomy Reports, 44, 790

\bibitem[\protect\citeauthoryear{{Mel{\'e}ndez}, {Asplund}, {Alves-Brito},
  {Cunha}, {Barbuy}, {Bessell}, {Chiappini}, {Freeman}, {Ram{\'{\i}}rez},
  {Smith} \& {Yong}}{{Mel{\'e}ndez} et~al.}{2008}]{Melendez08}
{Mel{\'e}ndez} J.,  {Asplund} M.,  {Alves-Brito} A.,  {Cunha} K.,  {Barbuy} B.,
   {Bessell} M.~S.,  {Chiappini} C.,  {Freeman} K.~C.,  {Ram{\'{\i}}rez} I.,
  {Smith} V.~V.,    {Yong} D.,  2008, \aap, 484, L21

\bibitem[\protect\citeauthoryear{{Mermilliod}, {Mayor} \& {Udry}}{{Mermilliod}
  et~al.}{2008}]{MeMU08}
{Mermilliod} J.~C.,  {Mayor} M.,    {Udry} S.,  2008, \aap, 485, 303

\bibitem[\protect\citeauthoryear{{Michaud}, {Richard}, {Richer} \&
  {VandenBerg}}{{Michaud} et~al.}{2004}]{Mi04}
{Michaud} G.,  {Richard} O.,  {Richer} J.,    {VandenBerg} D.~A.,  2004, \apj,
  606, 452

\bibitem[\protect\citeauthoryear{{Michaud}, {Richer} \& {Richard}}{{Michaud}
  et~al.}{2010}]{Michaud10}
{Michaud} G.,  {Richer} J.,    {Richard} O.,  2010, \aap, 510, A104

\bibitem[\protect\citeauthoryear{{Mikolaitis}, {Tautvai{\v s}ien{\.e}},
  {Gratton}, {Bragaglia} \& {Carretta}}{{Mikolaitis}
  et~al.}{2010}]{Mikolaitis10}
{Mikolaitis} {\v S}.,  {Tautvai{\v s}ien{\.e}} G.,  {Gratton} R.,  {Bragaglia}
  A.,    {Carretta} E.,  2010, \mnras, 407, 1866

\bibitem[\protect\citeauthoryear{{Mikolaitis}, {Tautvai{\v s}ien{\.e}},
  {Gratton}, {Bragaglia} \& {Carretta}}{{Mikolaitis}
  et~al.}{2011}]{Mikolaitis11}
{Mikolaitis} {\v S}.,  {Tautvai{\v s}ien{\.e}} G.,  {Gratton} R.,  {Bragaglia}
  A.,    {Carretta} E.,  2011, \mnras, 413, 2199

\bibitem[\protect\citeauthoryear{{Mishenina}, {Bienaym\'e}, {Gorbaneva},
  {Charbonnel}, {Soubiran}, {Korotin} \& {Kovtyukh}}{{Mishenina}
  et~al.}{2006}]{Mi06}
{Mishenina} T.~V.,  {Bienaym\'e} O.,  {Gorbaneva} T.~I.,  {Charbonnel} C.,
  {Soubiran} C.,  {Korotin} S.~A.,    {Kovtyukh} V.~V.,  2006, \aap, 456, 1109

\bibitem[\protect\citeauthoryear{{Montalb{\'a}n} \&
  {Schatzman}}{{Montalb{\'a}n} \& {Schatzman}}{2000}]{MoSc00}
{Montalb{\'a}n} J.,  {Schatzman} E.,  2000, \aap, 354, 943

\bibitem[\protect\citeauthoryear{{Mowlavi}}{{Mowlavi}}{1999}]{Mow99}
{Mowlavi} N.,  1999, \aap, 350, 73

\bibitem[\protect\citeauthoryear{{Palacios}, {Charbonnel}, {Talon} \&
  {Siess}}{{Palacios} et~al.}{2006}]{PC06}
{Palacios} A.,  {Charbonnel} C.,  {Talon} S.,    {Siess} L.,  2006, \aap, 453,
  261

\bibitem[\protect\citeauthoryear{{Palmerini}, {La Cognata}, {Cristallo} \&
  {Busso}}{{Palmerini} et~al.}{2011}]{Palmerini11}
{Palmerini} S.,  {La Cognata} M.,  {Cristallo} S.,    {Busso} M.,  2011, \apj,
  729, 3

\bibitem[\protect\citeauthoryear{{Pancino}, {Carrera}, {Rossetti} \&
  {Gallart}}{{Pancino} et~al.}{2010}]{Pancino10}
{Pancino} E.,  {Carrera} R.,  {Rossetti} E.,    {Gallart} C.,  2010, \aap, 511,
  A56+

\bibitem[\protect\citeauthoryear{{Pasquini}, {Randich}, {Zoccali}, {Hill},
  {Charbonnel} \& {Nordstr{\"o}m}}{{Pasquini} et~al.}{2004}]{PRZ04}
{Pasquini} L.,  {Randich} S.,  {Zoccali} M.,  {Hill} V.,  {Charbonnel} C.,
  {Nordstr{\"o}m} B.,  2004, \aap, 424, 951

\bibitem[\protect\citeauthoryear{{Paulson}, {Sneden} \& {Cochran}}{{Paulson}
  et~al.}{2003}]{Paulson03}
{Paulson} D.~B.,  {Sneden} C.,    {Cochran} W.~D.,  2003, \aj, 125, 3185

\bibitem[\protect\citeauthoryear{{Pavlenko}, {Jones} \& {Longmore}}{{Pavlenko}
  et~al.}{2003}]{P03}
{Pavlenko} Y.~V.,  {Jones} H.~R.~A.,    {Longmore} A.~J.,  2003, \mnras, 345,
  311

\bibitem[\protect\citeauthoryear{{Perryman}, {Brown}, {Lebreton}, {Gomez},
  {Turon}, {de Strobel}, {Mermilliod}, {Robichon}, {Kovalevsky} \&
  {Crifo}}{{Perryman} et~al.}{1998}]{Perryman98}
{Perryman} M.~A.~C.,  {Brown} A.~G.~A.,  {Lebreton} Y.,  {Gomez} A.,  {Turon}
  C.,  {de Strobel} G.~C.,  {Mermilliod} J.~C.,  {Robichon} N.,  {Kovalevsky}
  J.,    {Crifo} F.,  1998, \aap, 331, 81

\bibitem[\protect\citeauthoryear{{Pilachowski}, {Sneden}, {Freeland} \&
  {Casperson}}{{Pilachowski} et~al.}{2003}]{Pi03}
{Pilachowski} C.,  {Sneden} C.,  {Freeland} E.,    {Casperson} J.,  2003, \aj,
  125, 794

\bibitem[\protect\citeauthoryear{{Pinsonneault}}{{Pinsonneault}}{1997}]{Pin97}
{Pinsonneault} M.,  1997, \araa, 35, 557

\bibitem[\protect\citeauthoryear{{Ralchenko}, {Kramida}, {Reader} \& {NIST ASD
  Team}}{{Ralchenko} et~al.}{2010}]{NIST10}
{Ralchenko} Y.,  {Kramida} A.,  {Reader} J.,    {NIST ASD Team} 2010, {NIST
  Atomic Database (version 4.0)}.
National Institute of Standards and Technology, http://physics.nist.gov/asd

\bibitem[\protect\citeauthoryear{{Randich}, {Sestito}, {Primas}, {Pallavicini}
  \& {Pasquini}}{{Randich} et~al.}{2006}]{R06}
{Randich} S.,  {Sestito} P.,  {Primas} F.,  {Pallavicini} R.,    {Pasquini} L.,
   2006, \aap, 450, 557

\bibitem[\protect\citeauthoryear{{Recio-Blanco} \& {de Laverny}}{{Recio-Blanco}
  \& {de Laverny}}{2007}]{RBdL07}
{Recio-Blanco} A.,  {de Laverny} P.,  2007, \aap, 461, L13

\bibitem[\protect\citeauthoryear{{Sansonetti} \& {Martin}}{{Sansonetti} \&
  {Martin}}{2005}]{SansonettiMartin05}
{Sansonetti} J.~E.,  {Martin} W.~C.,  2005, J. Phys. Chem. ref. Data, 34, 559+

\bibitem[\protect\citeauthoryear{{Sato}, {Izumiura}, {Toyota}, {Kambe},
  {Takeda}, {Masuda}, {Omiya}, {Murata}, {Itoh}, {Ando}, {Yoshida}, {Ikoma},
  {Kokubo} \& {Ida}}{{Sato} et~al.}{2007}]{Sato07}
{Sato} B.,  {Izumiura} H.,  {Toyota} E.,  {Kambe} E.,  {Takeda} Y.,  {Masuda}
  S.,  {Omiya} M.,  {Murata} D.,  {Itoh} Y.,  {Ando} H.,  {Yoshida} M.,
  {Ikoma} M.,  {Kokubo} E.,    {Ida} S.,  2007, \apj, 661, 527

\bibitem[\protect\citeauthoryear{{Sbordone}}{{Sbordone}}{2005}]{Sbordone05}
{Sbordone} L.,  2005, Memorie della Societa Astronomica Italiana Supplement, 8,
  61

\bibitem[\protect\citeauthoryear{{Sbordone}, {Bonifacio}, {Castelli} \&
  {Kurucz}}{{Sbordone} et~al.}{2004}]{Sbordone04}
{Sbordone} L.,  {Bonifacio} P.,  {Castelli} F.,    {Kurucz} R.~L.,  2004,
  Memorie della Societa Astronomica Italiana Supplement, 5, 93

\bibitem[\protect\citeauthoryear{{Schuler}, {Hatzes}, {King}, {K{\"u}rster} \&
  {The}}{{Schuler} et~al.}{2006}]{Schu06}
{Schuler} S.~C.,  {Hatzes} A.~P.,  {King} J.~R.,  {K{\"u}rster} M.,    {The}
  L.-S.,  2006, \aj, 131, 1057

\bibitem[\protect\citeauthoryear{{Schuler}, {King} \& {The}}{{Schuler}
  et~al.}{2009}]{Schuler09}
{Schuler} S.~C.,  {King} J.~R.,    {The} L.-S.,  2009, \apj, 701, 837

\bibitem[\protect\citeauthoryear{{Sestito}, {Randich} \& {Bragaglia}}{{Sestito}
  et~al.}{2007}]{Ses07}
{Sestito} P.,  {Randich} S.,    {Bragaglia} A.,  2007, \aap, 465, 185

\bibitem[\protect\citeauthoryear{{Shi}, {Gehren} \& {Zhao}}{{Shi}
  et~al.}{2004}]{SGZ04}
{Shi} J.~R.,  {Gehren} T.,    {Zhao} G.,  2004, \aap, 423, 683

\bibitem[\protect\citeauthoryear{{Smiljanic}, {Gauderon}, {North}, {Barbuy},
  {Charbonnel} \& {Mowlavi}}{{Smiljanic} et~al.}{2009}]{Sm09b}
{Smiljanic} R.,  {Gauderon} R.,  {North} P.,  {Barbuy} B.,  {Charbonnel} C.,
  {Mowlavi} N.,  2009, \aap, 502, 267

\bibitem[\protect\citeauthoryear{{Smiljanic}, {Pasquini}, {Charbonnel} \&
  {Lagarde}}{{Smiljanic} et~al.}{2010}]{Sm10}
{Smiljanic} R.,  {Pasquini} L.,  {Charbonnel} C.,    {Lagarde} N.,  2010, \aap,
  510, A50

\bibitem[\protect\citeauthoryear{{Smith}, {Hinkle}, {Cunha}, {Plez}, {Lambert},
  {Pilachowski}, {Barbuy}, {Mel{\'e}ndez}, {Balachandran}, {Bessell},
  {Geisler}, {Hesser} \& {Winge}}{{Smith} et~al.}{2002}]{Smith02}
{Smith} V.~V.,  {Hinkle} K.~H.,  {Cunha} K.,  {Plez} B.,  {Lambert} D.~L.,
  {Pilachowski} C.~A.,  {Barbuy} B.,  {Mel{\'e}ndez} J.,  {Balachandran} S.,
  {Bessell} M.~S.,  {Geisler} D.~P.,  {Hesser} J.~E.,    {Winge} C.,  2002,
  \aj, 124, 3241

\bibitem[\protect\citeauthoryear{{Sneden}, {Pilachowski} \&
  {Vandenberg}}{{Sneden} et~al.}{1986}]{SPVB86}
{Sneden} C.,  {Pilachowski} C.~A.,    {Vandenberg} D.~A.,  1986, \apj, 311, 826

\bibitem[\protect\citeauthoryear{{Soubiran} \& {Girard}}{{Soubiran} \&
  {Girard}}{2005}]{SG05}
{Soubiran} C.,  {Girard} P.,  2005, \aap, 438, 139

\bibitem[\protect\citeauthoryear{{Soubiran}, {Le Campion}, {Cayrel de Strobel}
  \& {Caillo}}{{Soubiran} et~al.}{2010}]{Soubiran10}
{Soubiran} C.,  {Le Campion} J.,  {Cayrel de Strobel} G.,    {Caillo} A.,
  2010, \aap, 515, A111+

\bibitem[\protect\citeauthoryear{{Spite}, {Cayrel}, {Hill}, {Spite}, {Fran\c
  cois}, {Plez}, {Bonifacio}, {Molaro}, {Depagne}, {Andersen}, {Barbuy},
  {Beers} \& {Nordstr\"om} B.~{Primas}}{{Spite} et~al.}{2006}]{Sp06}
{Spite} M.,  {Cayrel} R.,  {Hill} V.,  {Spite} F.,  {Fran\c cois} P.,  {Plez}
  B.,  {Bonifacio} P.,  {Molaro} P.,  {Depagne} E.,  {Andersen} J.,  {Barbuy}
  B.,  {Beers} T.~C.,    {Nordstr\"om} B.~{Primas} F.,  2006, \aap, 455, 291

\bibitem[\protect\citeauthoryear{{Suda}, {Yamada}, {Katsuta}, {Komiya},
  {Ishizuka}, {Aoki} \& {Fujimoto}}{{Suda} et~al.}{2011}]{Suda11}
{Suda} T.,  {Yamada} S.,  {Katsuta} Y.,  {Komiya} Y.,  {Ishizuka} C.,  {Aoki}
  W.,    {Fujimoto} M.~Y.,  2011, \mnras, 412, 843

\bibitem[\protect\citeauthoryear{{Takeda}, {Zhao}, {Takada-Hidai}, {Chen},
  {Saito} \& {Zhang}}{{Takeda} et~al.}{2003}]{Tak03}
{Takeda} Y.,  {Zhao} G.,  {Takada-Hidai} M.,  {Chen} Y.-Q.,  {Saito} Y.-J.,
  {Zhang} H.-W.,  2003, Chinese Journal of Astronomy and Astrophysics, 3, 316

\bibitem[\protect\citeauthoryear{{Talon} \& {Charbonnel}}{{Talon} \&
  {Charbonnel}}{2005}]{TC05}
{Talon} S.,  {Charbonnel} C.,  2005, \aap, 440, 981

\bibitem[\protect\citeauthoryear{{Tautvai{\v s}ien{\.e}}, {Edvardsson},
  {Puzeras}, {Barisevi{\v c}ius} \& {Ilyin}}{{Tautvai{\v s}ien{\.e}}
  et~al.}{2010}]{Tautvaisiene10}
{Tautvai{\v s}ien{\.e}} G.,  {Edvardsson} B.,  {Puzeras} E.,  {Barisevi{\v
  c}ius} G.,    {Ilyin} I.,  2010, \mnras, 409, 1213

\bibitem[\protect\citeauthoryear{{Tautvai{\v s}ien{\.e}}, {Edvardsson},
  {Puzeras} \& {Ilyin}}{{Tautvai{\v s}ien{\.e}} et~al.}{2005}]{T05}
{Tautvai{\v s}ien{\.e}} G.,  {Edvardsson} B.,  {Puzeras} E.,    {Ilyin} I.,
  2005, \aap, 431, 933

\bibitem[\protect\citeauthoryear{{Tautvai{\v s}ien{\.e}}, {Edvardsson},
  {Tuominen} \& {Ilyin}}{{Tautvai{\v s}ien{\.e}} et~al.}{2000}]{T00}
{Tautvai{\v s}ien{\.e}} G.,  {Edvardsson} B.,  {Tuominen} I.,    {Ilyin} I.,
  2000, \aap, 360, 499

\bibitem[\protect\citeauthoryear{{van Leeuwen}}{{van
  Leeuwen}}{2009}]{VanLeeuwen09}
{van Leeuwen} F.,  2009, \aap, 497, 209

\bibitem[\protect\citeauthoryear{{Young}, {Knierman}, {Rigby} \&
  {Arnett}}{{Young} et~al.}{2003}]{Young03}
{Young} P.~A.,  {Knierman} K.~A.,  {Rigby} J.~R.,    {Arnett} D.,  2003, \apj,
  595, 1114

\end{thebibliography}

\end{document}